\documentclass[twocolumn,showpacs,superscriptaddress,preprintnumbers,amsmath,amssymb]{revtex4}
\usepackage{graphicx}
\usepackage{dcolumn}
\usepackage{bm}

\newcommand \beq{\begin{equation}}
\newcommand \eeq{\end{equation}}

\def\simge{\mathrel{%
     \rlap{\raise 0.511ex \hbox{$>$}}{\lower 0.511ex \hbox{$\sim$}}}}
\def\simle{\mathrel{
     \rlap{\raise 0.511ex \hbox{$<$}}{\lower 0.511ex \hbox{$\sim$}}}}

\begin{document}
\preprint{ECT*-05-06; CERN-PH-TH-2005-143; SPhT-T05/126; TIFR/TH/05-30}
\title{Elliptic flow and incomplete equilibration at RHIC}
\author{Rajeev S. Bhalerao}
\affiliation{Department of Theoretical Physics, TIFR,
   Homi Bhabha Road, Colaba, Mumbai 400 005, India}
\author{Jean-Paul Blaizot}
\affiliation{ECT*, Villa Tambosi, strada delle Tabarelle 286, I38050
   Villazzano (TN), Italy}
\author{Nicolas Borghini}
\affiliation{Physics Department, Theory Division, CERN, CH-1211 Geneva 23,
   Switzerland}
\author{Jean-Yves Ollitrault}
\affiliation{Service de Physique Th\'eorique, CEA/DSM/SPhT, Unit\'e de
recherche associ\'ee au CNRS,\\ F-91191 Gif-sur-Yvette Cedex, France.}

\date{\today}

\begin{abstract}
We argue that RHIC data, in particular those on the anisotropic flow
coefficients $v_2$ and $v_4$, suggest that the matter produced in the early
stages of nucleus-nucleus collisions is incompletely thermalized.
We interpret the parameter $(1/S)(dN/dy)$, where $S$
is the transverse area of the collision zone and $dN/dy$ the multiplicity
density, as an indicator of the number of collisions per particle at the time
when elliptic flow is established, and hence as a  measure of  the 
degree of equilibration.
This number serves as a control parameter which can be varied
experimentally by changing the system size, the centrality or the beam energy.
We provide predictions for Cu--Cu collisions at RHIC as well as for
Pb--Pb collisions at the LHC.
\end{abstract}

\pacs{25.75.Ld, 24.10.Nz}

\maketitle

The observation of azimuthal anisotropy in the production of particles in
ultra-relativistic nucleus-nucleus collisions, especially the 
so-called elliptic
flow $v_2$~\cite{Ollitrault:1998vz}, is one of the highlights of the RHIC heavy
ion program~\cite{Adams:2004bi,Adler:2003kt}.
The phenomenon, first identified in this regime at the CERN
SPS~\cite{Alt:2003ab}, reflects the anisotropy of the region of overlap
of the nuclei and is a direct consequence of the reinteractions 
between the produced particles.
In the limit where the collisions are frequent enough to drive the 
system quickly
to local equilibrium, fluid dynamics provides an intuitive physical explanation
for the origin of anisotropic flow: particles tend to go in the direction of
the strongest pressure gradients, hence preferably in the collision
plane~\cite{Ollitrault:1992bk}.
Of course, local equilibrium is not a necessary condition for elliptic flow,
but it is commonly accepted that deviation from equilibrium can only reduce the
magnitude of the effect.

In fact, the main characteristics of the observed elliptic flow are reasonably
well described by ideal fluid dynamics~\cite{Kolb:2003dz}, while requiring
unreasonably large cross sections in transport
models~\cite{Molnar:2001ux,Molnar:2004yh}.
The ability of the former to reproduce both the elliptic flow and 
single-particle spectra for measured hadrons with $p_T\simle 2$~GeV/$c$ near
midrapidity in minimum-bias collisions is considered
a significant
finding at RHIC; by contrast, at the SPS a simultaneous fit of both
observables appears impossible.
The dependence of the flow pattern on hadron masses further supports the
hydrodynamical picture.
Very strong conclusions have been drawn from this apparent
success~\cite{Teaney:2001av}:
one argues that local equilibrium has to be 
established on short time scales ($\sim 0.6$~fm/$c$~\cite{Kolb:2003dz}) while
viscosity should be negligible, suggesting a ``perfect fluid'' behaviour for
the created matter~\cite{Gyulassy:2004zy}.

In this Letter we would like to question these conclusions.
On the one hand, the short time scale for equilibration is difficult to account
for microscopically~\cite{Baier:2000sb} (although  it has been
   argued recently  that plasma instabilities may
  provide a mechanism for fast thermalization; see, e.g.,
Refs.~\cite{Arnold:2004ti,Rebhan:2004ur}).
On the other hand, several features of the data clearly signal the breakdown of
the hydrodynamical description and are more naturally understood  if 
the constraint of
local equilibrium is relaxed.
Thus, the point of view that we shall adopt here is that 
matter produced at RHIC is not fully
equilibrated, and we shall explore the consequences of such an 
assumption. As we shall see, this leads to simple predictions   that 
can be easily tested. 

This Letter is organized as follows:
We first recall the  essential features of the elliptic flow using an
ideal hydrodynamic picture. We then show that incomplete 
thermalization leads to specific
deviations from hydrodynamical behaviour: these concern in particular
the dependence of moments $v_2$ and $v_4$ of the azimuthal 
distribution on the system size and the collision
centrality. Note that throughout this paper, we shall only consider 
the average values of $v_2$ and $v_4$ over all particles at a given 
rapidity; the effects of partial thermalization on the elliptic flow
of identified particles, in particular its $p_T$-dependence,
have been discussed in Ref.~\cite{Teaney:2003pb}.

Elliptic flow originates from the anisotropy of the initial
   matter distribution.
  In hydrodynamics, the dependences of the  elliptic flow on the 
system size and the centrality
are essentially determined
by the spatial eccentricity, defined as
\beq
\label{epsilon}
\epsilon=\frac{\langle y^2- x^2\rangle}{\langle y^2+ x^2\rangle},
\eeq
where  $x$ and $y$ are coordinates in the
plane perpendicular to the collision axis (with the  $x$-direction
in the collision plane, and the origin midway between the centers
of the two nuclei). Angular brackets $\langle \cdot\rangle$ denote an average
weighted with the initial entropy
density $s(x,y)$ (at a given rapidity).

Elliptic flow develops gradually in the system as it evolves.
If the speed of sound $c_s$ is constant, the 
natural time scale is of the order of $\bar R/c_s$, where $\bar R$ is 
a measure of the transverse size of the system: 
the anisotropy of the
momentum distributions can
  be fully achieved only once all parts of the system are ``informed"  about the
initial spatial anisotropy, and that takes a time of the order of 
$\bar R/c_s$. We define $\bar R$ through
$1/\bar R=\sqrt{1/\langle x^2\rangle+1/\langle y^2\rangle}$ (since flow is
an effect of pressure {\em gradients\/}, $\bar R$ is a more natural choice
than, e.g., the rms radius).

\begin{figure}
\includegraphics*[width=\linewidth]{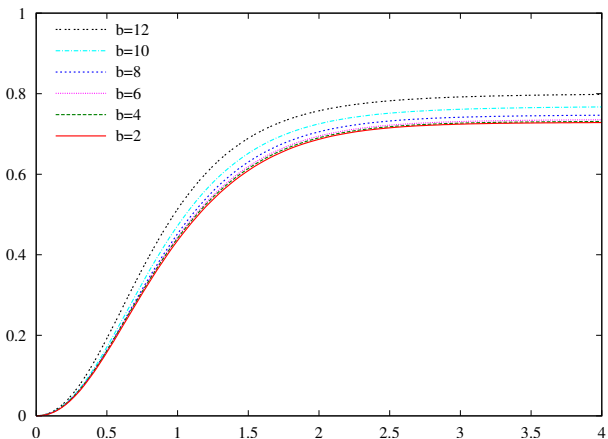}
\caption{(Color online) $v_2/\epsilon$ as a function of $c_s (t-t_0)/\bar R$
   for various impact parameter values of Au--Au collisions at
   $\sqrt{s_{NN}}=200$~GeV with $c_s=c/\sqrt{3}$.\label{fig:v2_vs_b}}
\end{figure}
  Hydrodynamical results are presented in Fig.~\ref{fig:v2_vs_b},
which displays the time evolution of elliptic flow
(i.e., the value of $v_2$ that one would
observe if the system was to decouple at time $t$)
for various centralities.
The procedure is detailed in Ref.~\cite{Ollitrault:1992bk}. There,
$v_2$ is  defined as
$v_2\equiv\langle p_x^2-p_y^2\rangle/\langle p_x^2+p_y^2\rangle$,
where average values are taken over all particles.
This
yields values of $v_2$ that are typically a factor 2 larger
than those obtained with the more conventional definition,
$v_2\equiv\langle (p_x^2-p_y^2)/(p_x^2+p_y^2)\rangle$.
  The longitudinal expansion is assumed boost-invariant;
the hydrodynamical evolution starts at a time
$t_0=0.6$~fm/$c$ after the collision. We checked that the
results are independent of the value of $t_0$
as long as $t_0$ is much smaller than $\bar R/c_s$.
In line with the discussion above, we have plotted  $v_2$ divided  by 
$\epsilon$,
and the elapsed time $t-t_0$  divided by the characteristic time   
$\bar R/c_s$.
This results in
an almost perfect scaling for a large range of
impact parameters (from $b=2$ to $b=12$~fm) for which the eccentricity varies
by more than one order of magnitude and $\bar R$ by a factor 2 (see
Table~\ref{tab:Au-Au} below).
Note, in particular,  that 
the final value of $v_2$ is independent 
of the system size ($\bar R$) for a given shape ($\epsilon$). 
This is a consequence of the scale invariance of ideal fluid dynamics.

\begin{figure}
\includegraphics*[width=\linewidth]{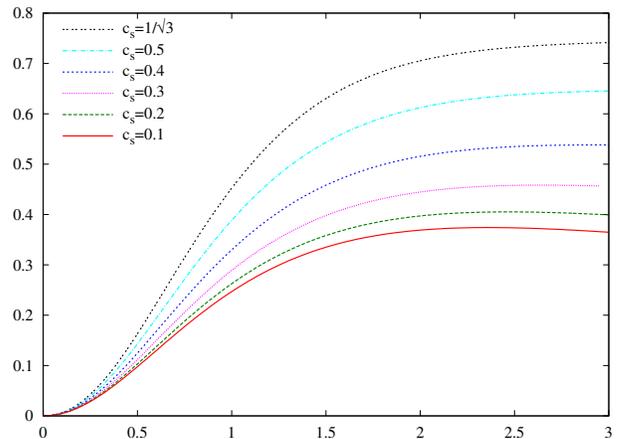}
\caption{(Color online) $v_2/\epsilon$ as a function of $c_s (t-t_0)/\bar R$
   for various values of $c_s$ in the case of  Au--Au collisions at
   $\sqrt{s_{NN}}=200$~GeV with impact parameter $b=8$~fm.
   \label{fig:eos1b}}
\end{figure}

The magnitude of
   $v_2$ also depends on the fluid properties, in particular
on the speed of sound $c_s$. In order to illustrate this dependence,
we have repeated the calculation for various values of
$c_s$.
The results are presented in Fig.~\ref{fig:eos1b}.
When $c_s$ is large enough ($c_s\simge 0.3$), $v_2$ is proportional to
$c_s$.
This no longer holds when $c_s$ becomes small ($c_s\simle 0.2$): in this
``nonrelativistic'' regime, the expansion of the system is entirely controlled
by the dimensionless parameter $c_s(t-t_0)/\bar R$:  we have
checked that  $v_2/\epsilon$ becomes a universal curve for $c_s<0.1$.
Note that the results plotted in Fig.~\ref{fig:eos1b} have been 
obtained by assuming that $c_s$ is constant throughout
the evolution.
In the real world, $c_s$ varies.
In particular, if the system enters a regime where the equation of state is
soft, i.e., where the speed of sound is small, the resulting $v_2$ may be
significantly reduced~\cite{Huovinen:2005gy}.

In summary, we have seen that within ideal fluid dynamics the
   final value of the elliptic flow $v_2$ is proportional to the initial
spatial eccentricity
$\epsilon$~\cite{Ollitrault:1992bk},  is independent of the system
   size $\bar R$,
grows with the speed of sound
$c_s$,
and has a characteristic build-up time $\sim\bar R/c_s$.
Note that this time scale for the build-up of the elliptic flow is of the same
   order of magnitude (although somewhat larger)
in transport calculations (see, e.g., Refs.~\cite{Molnar:2004yh,Chen:2005mr}).

We start now  exploring  the consequences of incomplete 
equilibration. To do so,  we
characterize the degree of thermalization by a dimensionless
parameter,  the Knudsen number $K$~\cite{Knudsen}. By definition, $K^{-1}$ is
the typical number of collisions per particle. Local thermal
equilibrium is achieved if $K^{-1}\gg 1$.
We are going to show that $K^{-1}$ can be determined from the data, as it
is proportional to
$(1/S)\,dN/dy$, where $dN/dy$ is the total multiplicity
density, while $S\equiv2\pi\sqrt{\langle x^2\rangle\langle y^2\rangle}$ 
is a measure of the transverse area of the collision zone 
(with this definition of $S$, 
which is larger by a factor 2 than that adopted in Ref.~\cite{Alt:2003ab},
$(1/S)\,dN/dy$ is the maximum value of the density for a Gaussian density
distribution).

The relevant length and time scales
for elliptic flow are $\bar R$ and $\bar R/c_s$, respectively.
Therefore, the typical number of collisions per particle is given by $\bar
R/\lambda$, where $\lambda$ is the mean free path
at time $\bar R/c_s$. This mean free path $\lambda$ depends on the particle
density $n$: $1/\lambda=\sigma n$, with $\sigma$ a cross section
characterizing the interactions among the produced particles. The 
particle density, in turn, depends
on time.
We assume that the total particle number is conserved throughout the
evolution: this is justified by the observation that $n$ is
proportional to the entropy density, and that entropy is conserved. This
   allows us to estimate the particle number density
at time $\tau$ from the relation
\beq
\label{density}
c\tau n(\tau) \sim\frac{1}{S} \frac{dN}{dy},
\eeq
valid for times $\tau\simle \bar R/c_s$, i.e., as long as the
transverse size of the system does not vary significantly.
One eventually obtains
\beq
\label{knud}
K^{-1} = \frac{\bar R}{\lambda} = \bar R\,\sigma\, n\!\left( \frac{\bar 
R}{c_s}\right) =
\frac{\sigma}{S}\frac{dN}{dy}\frac{c_s}{c}.
\eeq

Tables~\ref{tab:Au-Au} and \ref{tab:Cu-Cu} provide
numerical values of $(1/S)\,dN/dy$
for two colliding systems, Au--Au and Cu--Cu, at the same
center-of-mass energy $\sqrt{s_{NN}}=200$~GeV.
They are given in mb$^{-1}$, making the conversion into a
Knudsen number easy once the cross section is known (up to 
a factor $c_s/c$).
With a typical partonic cross section of 3~mb, and a speed of sound
$c_s\sim c/\sqrt{3}$, one thus finds $K^{-1}\simeq 2$
for a semi-central Au--Au collision.
This is not much larger than unity, and the ideal fluid limit
probably requires larger cross sections, as already inferred from
transport calculations.
The Knudsen number can also be estimated from the
viscosity~\cite{Gupta:2005jk}, with similar conclusions.

\begin{table}[t]
\caption{Parameters for Au--Au collisions at $\sqrt{s_{NN}}=200$~GeV. See text
   for details.}
\label{tab:Au-Au}
\begin{tabular}{|r|r|c|r|c|c|}
\hline
  $b$ (fm) & $\epsilon$ & $\bar R$ (fm) & $\frac{dN}{dy}$ &
  $\frac{1}{S}\frac{dN}{dy}$ (mb$^{-1}$) &
  $n\!\left(\frac{\bar R}{c_s}\right)$ (fm$^{-3}$) \\ \hline
0&0& 2.07& 1050& 1.95& 5.4\\ \hline
2& 0.033& 2.02& 975& 1.90& 5.4\\ \hline
4& 0.115 & 1.89& 790& 1.77& 5.5\\ \hline
6& 0.215 & 1.68& 562& 1.55& 5.3\\ \hline
8& 0.315 & 1.45& 344& 1.23 & 4.9\\ \hline
10& 0.398 & 1.22 & 167 & 0.82 & 3.8\\ \hline
12& 0.433 & 1.04 & 55 & 0.37 & 2.0\\ \hline
\end{tabular}
\end{table}
\begin{table}
\caption{Parameters for Cu--Cu collisions at $\sqrt{s_{NN}}=200$~GeV}
\label{tab:Cu-Cu}
\begin{tabular}{|r|r|c|r|c|c|}
\hline
  $b$ (fm) & $\epsilon$ & $\bar R$ (fm) & $\frac{dN}{dy}$ &
  $\frac{1}{S}\frac{dN}{dy}$ (mb$^{-1}$) &
  $n\!\left(\frac{\bar R}{c_s}\right)$ (fm$^{-3}$) \\ \hline
0 & 0 & 1.42 & 275 &  1.09 & 4.5 \\ \hline
2 & 0.043 & 1.36 & 240 & 1.02 & 4.3 \\ \hline
4 & 0.141 & 1.23 & 159 & 0.83 & 3.9 \\ \hline
5.5 & 0.216 & 1.10 & 95 & 0.61 & 3.2 \\ \hline
6 & 0.237 & 1.06 & 77 & 0.53 & 2.9 \\ \hline
8 & 0.265 & 0.93 & 22 & 0.20 & 1.2\\ \hline
\end{tabular}
\end{table}

Incomplete equilibration breaks the scale invariance
of ideal fluid dynamics: $v_2$ depends on the system size $\bar R$ 
through $K^{-1}$. 
As $K^{-1}$, increases,
the magnitude of $v_2$ grows
linearly---the larger the number of collisions,
the larger the momentum anisotropy---,
eventually saturating when the system reaches
local equilibrium.
To fix ideas, one may have in mind the following simple formula,
which exhibits the correct qualitative behaviour:
\beq
\label{simpleformula}
v_2=v_2^{\rm hydro}\frac{K^{-1}}{K^{-1}+K^{-1}_0}\, ,
\eeq
where $v_2^{\rm hydro}$ is the value of $v_2$ obtained from 
hydrodynamics, i.e.,
corresponding to the large $K^{-1}$ limit, and
$K^{-1}_0$ is a number of order unity, whose  precise value  can only 
be determined through  an explicit transport
calculation. Referring to the   transport calculations in 
Ref.~\cite{Chen:2005mr},
one finds  only a modest increase of $v_2$ (by 40\%)
when $\sigma$ increases from 3 to 10~mb, 
$K^{-1}$ from  2 to 7, which corresponds to $K_0^{-1}\simeq 1.5$. 

We are now in position to examine whether the matter produced  at 
RHIC is in local equilibrium. To do so, we need to vary the 
value of $K^{-1}$.
This can be done by changing the system size $\bar R$, while keeping
the mean free path $\lambda$ constant [see Eq.~(\ref{knud})].
Since $\lambda$ depends on the density (explicitly, and through a 
potential dependence of the cross section on the density), this 
requires keeping the density constant.
Now, the last column of Table~\ref{tab:Au-Au} 
shows that the particle density remains approximately constant
in Au--Au collisions for a large range of impact parameters, 
between central events and collisions with $b=8$~fm. In this case,
$K^{-1}$
is then proportional to the system size $\bar R$
or, equivalently, to
$(1/S)\,dN/dy$ [see Eq.~(\ref{knud})].
For small $K^{-1}$, we expect from Eq.~(\ref{simpleformula}) that
$v_2/\epsilon$ is proportional to $K^{-1}$
(since $v_2^{\rm hydro}$ scales like $\epsilon$), i.e., that
$v_2/\epsilon$ scales like $(1/S)\, dN/dy$.
Indeed, such a linear variation was found in Ref.~\cite{Alt:2003ab} (see
Fig.~25), for both the RHIC Au--Au values and the SPS Pb--Pb values.
It even turns out that both linear dependences match well, although the
densities differ (the densities at SPS are typically 40\% smaller 
than at RHIC), which suggests that the cross section (and $c_s$) 
used in estimating the Knudsen number 
can be taken as roughly constant across the different beam
energies.

We take the above-mentioned plot in Ref.~\cite{Alt:2003ab} as evidence
that local equilibrium is not achieved, since $v_2/\epsilon$ is steadily
increasing with $(1/S)\,dN/dy$, without any hint at a saturation, for the whole
accessible experimental range.
A similar evidence comes from the variation of $v_2$ with pseudorapidity
$\eta$~\cite{Hirano:2001eu,Heinz:2004et}:
changing the latter induces a variation of the multiplicity density $dN/d\eta$,
hence of the control parameter at fixed geometry, which is reflected in
$v_2(\eta)$.
Finally, the decrease in the number of collisions as the transverse momentum
increases also results in a departure from equilibrium seen on $v_2(p_T)$
data~\cite{Teaney:2003pb}, and in transport calculations~\cite{Molnar:2004yh}.
In the latter, for the bulk of the particles ($p_T\lesssim 500$~MeV/$c$), $v_2$
already saturates for $\sigma$ of the order of 3~mb.
For higher values of $p_T$, the saturation occurs only above 10~mb.
Surprisingly, however, the saturation value is significantly below that given
by hydrodynamical computations, which calls for further investigation of the
discrepancy.

One can also study the effect of incomplete equilibration on
the fourth moment $v_4$. Repeating the same arguments as for
$v_2$, one expects that the behaviour of $v_4$ as a function of $K^{-1}$
is given, at least qualitatively, by an equation similar to
(\ref{simpleformula}).
 From the simple observation that both $v_2$ and $v_4$ are
 proportional to $K^{-1}$ for small $K^{-1}$, one expects that
$v_4/v_2^2$ {\em decreases\/} with $K^{-1}$,
reaching a minimum when the hydrodynamical regime is reached.
Since it was shown in Ref.~\cite{Borghini:2005kd} that ideal fluid dynamics
yields $v_4/v_2^2=1/2$ (a similar value was found in Ref.~\cite{Kolb:2003zi}
within a specific hydrodynamical model), 
one expects the ratio to be larger than $1/2$
if the system is not fully equilibrated, in agreement with the experimental
finding $v_4/v_2^2\sim 1.2$~\cite{Adams:2004bi}.

Further predictions can be made to test the assumption of 
incomplete thermalization.
In particular, studying smaller systems at the same center-of-mass
energy, where the density is roughly the same (compare 
Tables~\ref{tab:Au-Au} and \ref{tab:Cu-Cu}), one can obtain direct
information on the dependence of $v_2/\epsilon$ on the number of 
collisions per particle, $K^{-1}$. 
Consider, e.g., Cu--Cu collisions at $b=5.5$~fm, which corresponds to 
collisions
with the same centrality as Au--Au collisions with $b=8$~fm~\cite{Chen:2005mr}.
If the hydrodynamical regime were reached, $v_2/\epsilon$ would be roughly
independent of the system size: from the values of the eccentricities in
Table~\ref{tab:Cu-Cu}, one concludes that the values of elliptic flow in Au--Au
and Cu--Cu collisions would be then related by $v_2({\rm 
Cu})=0.69\,v_2({\rm Au})$.
If, on the contrary, we are  far from equilibrium,
so that $v_2/\epsilon$ is proportional to $K^{-1}$, instead of
constant, then the relationship reads $v_2({\rm Cu})=0.34\,v_2({\rm Au})$.
This prediction can be completed by an analogous one concerning $v_4/v_2^2$:
as argued before, this quantity is a decreasing function of the number of
collisions, hence should be larger in Cu--Cu collisions than it is in Au--Au
collisions.
Other  predictions can be made, regarding the future experiments at 
LHC: if, as we have argued,
the ideal-fluid limit is not  reached  even in the most central
collisions at RHIC, then $v_2/\epsilon$ should further increase in Pb--Pb
collisions at $\sqrt{s_{NN}}=5.5$~TeV.
In parallel, one can expect that $v_4/v_2^2$ will decrease, coming closer to
$1/2$.

The case for early thermalization at RHIC rests on the argument that
the $v_2$ measurements saturate the hydrodynamical limit. 
However, the latter may well be underestimated. 
Indeed, the pion data presented in Fig.~36 of Ref.~\cite{Adams:2004bi} 
suggest that the measured $v_2$ in central collisions {\em overshoots\/} 
the value obtained by a hydrodynamical computation,
$v_2({\rm data})>v_2({\rm hydro})$.
We know from the above discussion that it is possible to 
increase the hydrodynamical prediction for $v_2$ by increasing  the 
speed of sound.
Now, in present hydrodynamical calculations, increasing $c_s$ means taking a
harder equation of state, and this   spoils  the agreement with 
experimental momentum
spectra, which require a soft equation of state~\cite{Kolb:2003dz}. 
It is important to realize that this requirement is closely linked 
to the assumption that the system is fully thermalized 
(see \cite{Heinz:2002rs} for an alternative to 3-dimensional 
thermalization)
and, even more importantly, reaches chemical equilibrium: 
in a system in chemical equilibrium, there 
exists a one-to-one relationship between energy per particle (probed by
momentum spectra) and density (probed by $dN/dy$). 
Although fits to particle ratios by thermal models 
are usually interpreted as evidence for chemical
equilibrium, there is no direct experimental evidence that the particle
density in the system obeys the laws of thermodynamics. 
In fact, as we have seen before, 
data indicate that {\em kinetic\/} equilibrium is not attained at
RHIC, which in turn suggests that chemical equilibrium is not attained 
either.
(Even recent hydro calculations, which assume early chemical
freeze-out, start with a system in equilibrium~\cite{Hirano:2005wx}.)
If one drops the assumption of chemical equilibrium,
energy per particle and density become independent variables and 
the transverse momentum spectra no longer constrain the equation of
state. 

To summarize, relaxing the constraint  of chemical
  equilibrium allows for a natural explanation of RHIC data on
  elliptic flow. Deviations from local equilibrium lead to a 
characteristic dependence of observables such as $v_2/\epsilon$ and 
$v_4/v_2^2$ on the number of collisions, and this can be tested
experimentally. 

\section*{Acknowledgments}

We acknowledge the financial support from CEFIPRA, New Delhi, 
under its project no. 3104-3.

\end{document}